\documentclass[runningheads,a4paper]{llncsModified}
\usepackage{graphicx}
\usepackage{epstopdf}

\long\def\comment#1{}

\newlength{\figtblfootnotemargin}
\newlength{\figtblfootnotewidth}

\newcommand{\outcomment}[1]{
    \comment{
        #1
    }
}

\newcommand{\SquareOfSize}[2]{
 \fbox{\hsize #1cm \hbox to #1cm{\vbox{#2}}}
}

\newsavebox{\wholeWidthLine}
\sbox{\wholeWidthLine} {\rule[0.1in]{\textwidth}{.01in}}

\newcommand{\bq}{\begin{quote}}
\newcommand{\eq}{\end{quote}}
\newcommand{\be}{\begin{enumerate}}
\newcommand{\ee}{\end{enumerate}}
\newcommand{\bi}{\begin{itemize}}
\newcommand{\ei}{\end{itemize}}
\newcommand{\bie}{\begin{itemize}\begin{enumerate}}
\newcommand{\eie}{\end{enumerate}\end{itemize}}
\newcommand{\ba}{\begin{array}}
\newcommand{\ea}{\end{array}}
\newcommand{\btbl}{\begin{tabular}}
\newcommand{\etbl}{\end{tabular}}
\newcommand{\bequ}{\begin{displaymath}}
\newcommand{\eequ}{\end{displaymath}}
\newcommand{\bequa}{\begin{eqnarray*}}
\newcommand{\eequa}{\end{eqnarray*}}
\newcommand{\bc}{\begin{center}}
\newcommand{\ec}{\end{center}}
\newcommand{\btab}{\begin{tabbing}}
\newcommand{\etab}{\end{tabbing}}

\newcommand{\godown}{ \vspace*{0.3cm}}

\newcommand{\mybox}[1]{\mbox{\letterspace #1 \letterspace }}

\def\mpr#1{\ifmmode #1 \else #1 \fi}

\newcommand{\letterspace}{\hspace*{1mm}}

\newcommand{\DPATTRNAME}{{\large{{\bf SC}$_{attr}$}}}
\newcommand{\DPISANAME}{{\large{{\bf SC}$_{isA}$}}}
\newcommand{\DPINNAME}{{\large{{\bf SC}$_{in}$}}}

\newcommand{\sys}[0]{{\tt {\small MovieSim}}}

\begin{document}

\title{Similarity-based Browsing over \\ Linked Open Data}


\author{
        Michael Hickson$^{1}$,  \ Yannis Kargakis$^{1}$ \ and \ Yannis Tzitzikas$^{1,2}$
}

\institute{
 Department of Computer Science, University of Crete, Greece \\
 \and
  Institute of Computer Science,  FORTH-ICS, Greece \\
 Email: {$\{$hickson,kargakis,tzitzik$\}$@csd.uoc.gr}
}


\maketitle
\begin{abstract}
An increasing amount of data is published on the Web according
to the Linked Open Data (LOD) principles.
End users would like to browse these data in a flexible manner.
In this paper we focus
on similarity-based browsing
and we introduce a novel method
for computing the similarity between
two entities of a given RDF/S graph.
The distinctive characteristics
of the proposed metric
is that it is generic
(it can be used to compare nodes of any kind),
it takes into account the neighborhoods of the nodes, and
it is configurable
(with respect to the accuracy vs computational complexity tradeoff).
We demonstrate the behavior of the metric
using examples from an application over LOD.
Finally, we generalize and elaborate on
implementation approaches
harmonized with the
distributed nature of LOD
which can be used for computing the most similar entities
using neighborhood-based similarity metrics.
\end{abstract}




\

\section{Introduction}

The last years a vast amount of structured data has been published
as {\em Linked Open Data (LOD)}.
However, in their current form, they cannot be directly exploited by end users, since
better linking, browsing, presentation is required
({\em interaction} and {\em interfaces}
is one of the main research challenges of LOD
according to \cite{bizer2009linked}).
Our objective is to investigate
generic methods
for browsing and exploring  such data sets.
\comment{
        The vast and rapid development of the Semantic Web  has created new
        demands on the field of graph processing, and more particularly for
        RDF/S graphs.
        The computation of similarity between two entities of  a given graph
        is a matter that concerns the research community of
        information systems and  many approaches have
        been developed.
        \footnote{
            {\em
            to check that holds:
            most of them do not take into account the complexity of the
            relationships between the nodes that are compared and seem to lack
            the ability of finding similarities between nodes of different
            types.
            }
        }
}
\comment{
    Based on the pattern of these techniques we analyze and
    present a new similarity function, which calculates the similarity
    between two nodes of a graph, solving the issues mentioned above and
    we finally explain the results obtained from its execution by
    demonstrating an example.
}
Context and motivation for our work  was the design and
development of an online movie exploration system based on
Semantic Web technologies,
whose data are fetched dynamically from the LOD cloud,
and
offers
{\em similarity-based browsing}
for bypassing the need for query formulation
by end users.

In this paper, we motivate the need for
similarity-based browsing,
we identify  related requirements,
and we introduce
a new similarity function
for tackling them.
In brief the proposed similarity between two RDF nodes is actually the
Jaccard similarity coefficient evaluated
over the nodes of the
extended
(radius bounded)
neighborhoods
(containing both instance and schema nodes)
of the compared nodes.
A distinctive characteristic of
this metric is that each node that participates to
an intersection or union operation
of the Jaccard similarity coefficient,
is weighted by a value
based on its path distance from the compared nodes,
for promoting close matches over distant ones.
\comment{
    Therefore a broad scope of the
    functions applicability is its embedment in systems in order to
    provide recommendations, which would be provided by the similarities
    that emerge through appropriate graphs.

    The demonstration and
    explanation of our method is based on an ontology sample, which
    refers to a system with a knowledge base associated with movies.
}
In a nutshell, the distinctive characteristics of the proposed similarity metric is that:
 (a) it is type independent (it can compute similarity between any pair of resources),
 (b) it can be applied within a single KB
    (thus different from the methods which have been proposed
    for ontology matching), and
 (c) it offers to the designer
    (or end user) the flexibility to choose the
    appropriate depth
    depending on his needs (on accuracy or computational complexity).
Subsequently,
we describe implementation approaches
for
computing the most similar entities
and we analyze
implementation approaches
which are harmonized with the
distributed nature of LOD.
In particular
we show how a similarity function
can be {\em reversed}
for enabling the computation of similar pages
over the LOD without having to access the entire corpus.
Such methods can be used not only for the introduced similarity metric,
but for neighborhood-based similarity metrics in general.

The rest of this paper is organized as follows.
Section \ref{sec:AC} describes the  motivation and application context of our work.
Section \ref{sec:RW} discusses  related works.
Section \ref{sec:Notations} introduces the least number of symbols and notations required for
defining the similarity function.
Section \ref{sec:Similarity}  introduces the similarity function
and
Section \ref{sec:Evaluation}
demonstrates its  merits  over the running example.
Section \ref{sec:Implementation}
discusses implementation approaches
and shows how a similarity function
can be {\em reversed}.
Finally, Section \ref{sec:Concl} concludes the paper
and identifies issues for further research.

\section{Application Context}
\label{sec:AC}

The context of our work is an application over
the {\em Linked Open Data (LOD)}  cloud.
\comment{
    The last years a vast amount of structured data has been published
    as LOD.
    However, in their current form, they cannot be directly exploited by end users, since
    better linking, browsing, presentation is required.
}
Our objective was to design and develop a system which allows the flexible exploration
of {\em movie} information, based on information fetched from the
LOD cloud. The distinctive characteristics of this system,
called  \sys, are:
\begin{itemize}
    \item All information is fetched from the LOD cloud.
          This not only automates information updating,
          but enables the  application to provide
          always up-to-date information.
    \item It links the available in the LOD structured information,
         and enriches it with links to
         external information
         (plain Web pages).
\end{itemize}

\comment{
    In particular, all information is fetched dynamically from multiple accessible
    SPARQL  endpoints and
    from FreeBase's\footnote{http://www.freebase.com/}
    data cloud, using the provided API,
    allowing us to fetch structured data,
    over the HTTP
    protocol, in JSON and RDF format respectively.
}
Specifically from LinkedMDB\footnote{
    http://www.linkedmdb.org/
} the data are fetched in RDF format,
from its available SPARQL Endpoint,
while from Freebase\footnote{
    http://www.freebase.com/
}
data cloud
the data is fetched in JSON format through its provided API.
Regarding the linking of the data extracted from each source
we did not face any difficulty, since LinkedMDB
provides for each of its entities
a Unique Identifier,
through FreeBase's link, that represents it in Freebase's data cloud.

Since  most end users do not  have the technical  knowledge
(or the willingness)  to
formulate explicit SPARQL queries,
\sys\  provides a more user
friendly interaction,
namely
(a) keyword-based retrieval and
(b) similarity-based browsing.

To support
{\em keyword-based retrieval}
\sys\ periodically fetches information from LinkedMdb
and indexes it with the help of LARQ (Lucene+ARQ)\footnote{
    http://jena.sourceforge.net/ARQ/lucene-arq.html
}.
The availability of an index makes the evaluation of keyword queries very  fast.
We will not describe this functionality in detail
since keyword searching over structured data
is not the focus of this paper.


{\em Similarity-based browsing}
aims
at allowing users to explore the available information
without having to formulate structured queries.
Note that similarity-based browsing
is mainly offered for browsing
image and video databases (e.g. \cite{borth2008navidgator}),
but (to the best of our knowledge) has not been
applied over  RDF data.
\comment{
    In brief, our system shows information about
    {\tt Movies},
    {\tt Actors},
    {\tt Directors}
    and many other movie related types.
}

Regarding the presentation of information,
\sys\
supports various kinds of Web pages,
each one having a different role.
Keyword search is supported through a search box,
while the results of the query are viewed by a different kind of page.
The essential  category of
pages contains page types  for showing information about:
\bi
\item actors,
\item directors,
\item editors,
\item movies, and
\item writers.
\ei

Each
page type
presents information
which is dynamically fetched
and linked.
In addition,
the system provides a general purpose page type
to show information about entity types
that do not fall in one of the previous categories.
Below we present the information that we fetch for each
supported type, from each individual source.

\bc
{\footnotesize
\btbl{|l|l|}\hline
\multicolumn{2}{|c|}{\bf Movie}\\\hline
attribute & source \\\hline\hline
Title  & LinkedMDB \\\hline
Runtime &  LinkedMDB \\\hline
Initial Release Date & LinkedMDB \\\hline
Movie Actors & LinkedMDB \\\hline
Movie Writers &  LinkedMDB \\\hline
Movie Directors & LinkedMDB \\\hline
Movie Editors & LinkedMDB \\\hline
Image & Freebase \\\hline
Abstract &  Freebase \\\hline
Rating & Freebase \\\hline
Tagline & Freebase \\\hline
Genres & Freebase \\\hline
\etbl
\hspace*{.5cm}
\btbl{|l|l|}\hline
\multicolumn{2}{|c|}{\bf Actor}\\\hline
attribute & source \\\hline\hline
Actor Name & LinkedMDB \\\hline
Films Acted & LinkedMDB \\\hline
Image & Freebase \\\hline
Abstract & Freebase \\\hline
Birth Date & Freebase \\\hline
Birth Place & Freebase \\\hline
Nationality & Freebase \\\hline
\etbl
\hspace*{.5cm}
\btbl{|l|l|}\hline
\multicolumn{2}{|c|}{\bf Director}\\\hline
Director Name & LinkedMDB \\\hline
Films Directed & LinkedMDB \\\hline
Image & Freebase \\\hline
Abstract & Freebase \\\hline
Birth Date & Freebase \\\hline
Birth Place & Freebase \\\hline
Nationality &  Freebase \\\hline
\etbl

\btbl{|l|l|}\hline
\multicolumn{2}{|c|}{\bf Writer}\\\hline
Writer Name & LinkedMDB \\\hline
Films Writen & LinkedMDB \\\hline
Image & Freebase  \\\hline
Abstract & Freebase \\\hline
Birth Date & Freebase \\\hline
Birth Place & Freebase \\\hline
Nationality & Freebase \\\hline
\etbl
\hspace*{.5cm}
\btbl{|l|l|}\hline
\multicolumn{2}{|c|}{\bf Editor}\\\hline
Editor Name & LinkedMDB \\\hline
Films Edited  & LinkedMDB \\\hline
Image & Freebase \\\hline
Abstract & Freebase \\\hline
Birth Date & Freebase \\\hline
Birth Place & Freebase \\\hline
Nationality & Freebase \\\hline
\etbl
\hspace*{.5cm}
\btbl{|l|l|}\hline
\multicolumn{2}{|c|}{\bf General}\\\hline
Title & LinkedMDB \\\hline
Inbound Links & LinkedMDB \\\hline
Outbound Links & LinkedMDB \\\hline
Image & Freebase \\\hline
Abstract & Freebase \\\hline
\etbl
}
\ec

\comment{
    We should note, that part of the schema that we illustrate in our
    example is taken from the LinkedMDB data cloud and we have also added the classes level
    and some resources and properties in the instance level, in order to provide a more complete and
    accurate example.
    More specifically, we have added the following properties and resources: \\
    Properties: location, nationality, basedOn, writer \\
    Resources: DaVinciCode Book, Illuminati Book, Sherlock Holmes Book, Dan Brown, Conan Doyele
}

\begin{figure}[htbp]
    \centerline{\fbox{\includegraphics[height=76mm]{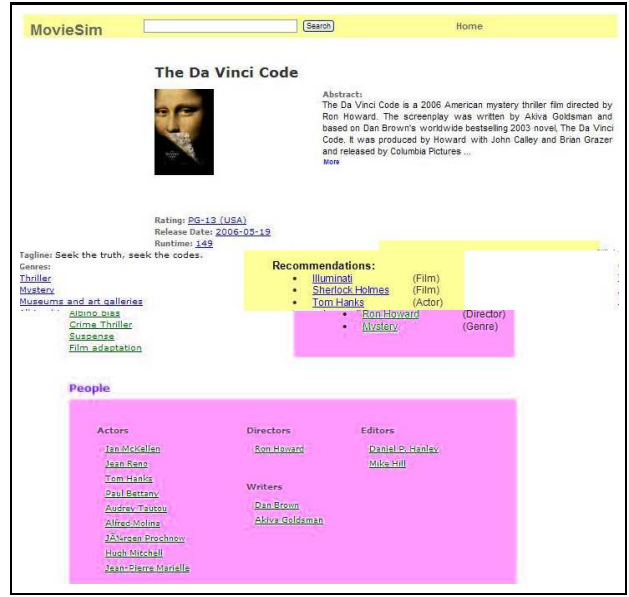}}}
    \caption{Movie Page}\label{fig:infopage}
\end{figure}

While the user views the page of one entity
he can continue browsing and exploring {\em similar entities}.
The similar entities are computed
using the similarity function that we will describe later on.
Since the similar entities can be numerous
and of different types,
only the entities with the highest similarity
should be suggested.
Figure \ref{fig:infopage} shows a screenshot of the  Web page
produced for the movie {\tt Da Vinci Code}.

Note that similarity-based browsing is actually
an alternative (essentially  complementary)
approach to the {\em facet-based browsing} \cite{Sacco09},
which is supported by systems like:
BrowseRdf \cite{oren06extending}, Humboldt, VisiNav \cite{Harth09}, Longwell \cite{Pietriga},
    Ontogator \cite{MakelaHS06}, /facet \cite{HildebrandOH06}, Camelis2 \cite{ferre2010conceptual}.
Facet-based browsing also bypasses the query formulation effort.
However, similarity-based browsing does not require from the user
to select the relationship through which two entities
are related. Instead,
the similarity value actually
quantifies several relationships (direct or path based)
and offers an aggregated form of relevance.

Similarity-based browsing can actually be offered
in the context of a facet-based browsing system.
Specifically,
a new facet can be defined
which shows the most similar entities.

\comment{NA TO SKEFTW KAI KATOPIN NA TO VALW:
    Less overloaded interface,
    the user may not necessarily want to decide exactly what is the relationship
    between the current and the next entity to view.
    Similarity quantifies several relationships (direct or path based),
    offering an aggregated form of relevance.
}

\begin{figure}[htbp]
    \centerline{\includegraphics[height=46mm]{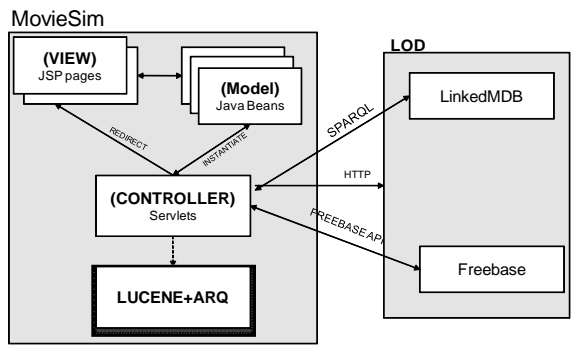}}
    \caption{The Architecture of \sys}\label{fig:mvc_figure}
\end{figure}

Figure \ref{fig:mvc_figure} sketches the architecture of \sys.
Its architecture is based on the {\em MVC} (Model View Controller) pattern,
meaning that all business logic is
implemented in Servlets and all  communication and data transfer
issues are dealt with the use of Java Beans
(one for each entity type mentioned earlier).
The presentation of data (page types)
is specified using  JSP pages
in order to separate the presentation design
from the application logic,
making easier the extension and modification of the system.

\section{Related Work on Similarity over RDF/S}
\label{sec:RW}

Since we focus on similarity-based browsing,
in this section we briefly review the related work that has been done.
In general, with the rapid development of the Semantic Web,
there has been an
increased interest in developing methods for finding similarities
between nodes in RDF/S graphs.
There are several related works mainly for the problem
of {\em ontology matching}.
Below we list and comment in brief the more related works.

\cite{thanh2005measuring}
presents a method for computing the similarity between two entities
coming from two different OWL DL ontologies.
The computation of similarity is
based on the extraction of information encoded in each entity's description.
The extracted components are then compared,
taking into account the predefined meanings of
OWL DL and RDF(S) primitives,
to produce partial component similarity values,
which are then combined
using predefined weights under a variable weighting scheme.

\cite{euzenat2004similarity} also proposes a similarity function
for entity matching between different OWL ontologies .

There are also algorithms (again for the problem of ontology matching)
which use the {\em edit} distance to find the lexical
similarity between two entities,
such as the MLMA+ algorithm \cite{alasoud2009empirical}
which,
amongst other measures, makes use of the Levenshtein (Edit)
distance \cite{levenshtein1966binary}.

Another algorithm (for ontology matching)
is presented in \cite{akbari2010novel}
for finding similarities
between two entities, of some given ontologies
based on the combination of structural and lexical information
provided by the ontology,
which is divided into three stages.
In the first
stage each entity is lexically analyzed, based on information given
from their labels and descriptions. The second stage involves the
comparison of the entities based on the structure of the graph,
while the third stage combines the results of the two previous
stages and produces a final result that represents the similarity
between the two entities.

Another related work
aiming at
identifying cases
where the same objects are identified by different URIs
in different datasets,
in the context of LOD,
is \cite{ObjectReconciliationESWC10}.

Finally,
\cite{schickel2007oss}
proposes a metric
for entity comparison in hierarchical ontologies
(however that work exploits only hierarchical relationships
and ignores properties).
\comment{
    We should stress that even in approaches where the similarity measure is applied on the
    same ontology, most of the times they are limited to a specific
    ontology structure.
    For instance,
    OSS \cite{schickel2007oss} proposes a metric
    for entity comparison in hierarchical ontologies
    (it exploits only hierarchical relationships,
    not properties).
}
\comment{
    This function
    computes the similarity between two entities,
    by inferring a score of one entity from the other
    and finding how much information
    is transferred between them.
    Finally it applies a function that
    converts the above score to a distance value.
    {\bf Sigoura agnoei tis properties?}.
}

Similar in spirit problem
is that of  {\em blank node matching}
which aims at defining a mapping
between the blank nodes
of two KBs
(related works include
PromptDiff \cite{Noy2002},  Ontoview \cite{klein2002ova},
CWM \cite{BernersLee2004},
RDFSync \cite{tummarello2007rdfsync}).

\godown

To synopsize,
most of the related works
aim at finding similarities between entities of {\em different}
knowledge bases.
Therefore
they mainly identify similarities
between entities of the {\em same type}.
Such approaches would not be convenient for our system,
since we would have to design several class-specific similarity functions,
i.e.
similarity functions
between movies and actors,
directors and actors,
writers and movies, and so on.
For this reason,
we decided to move towards
a similarity computation
method that  is {\em type-independent}
allowing the comparison
of entities of the same or different types.
At last
we should note
that the similarity function
that we needed for our system,
apart from being type-independent
should exploit both the instance and the schema layer
(for being able to compute similarities between entities which do not
belong to the same classes).

\comment{
    Instead according to our approach  the similarity between two
    entities is computed taking into account not only the correlations
    in the instance level but also in the schema (RDFS or OWL) level,
    where in
    each case the other level may or may not be taken into account,
    depending on the given parameters (depth).
}

\comment{
    To synopsize,
    from the aforementioned approaches, only the OSS \cite{schickel2007oss}
    can compute the  similarity between entities of the same ontology, but even
    in this case they emphasize on hierarchical ontologies.
}

\begin{figure*}[htb]
    \centerline{\fbox{\includegraphics[height=12cm]{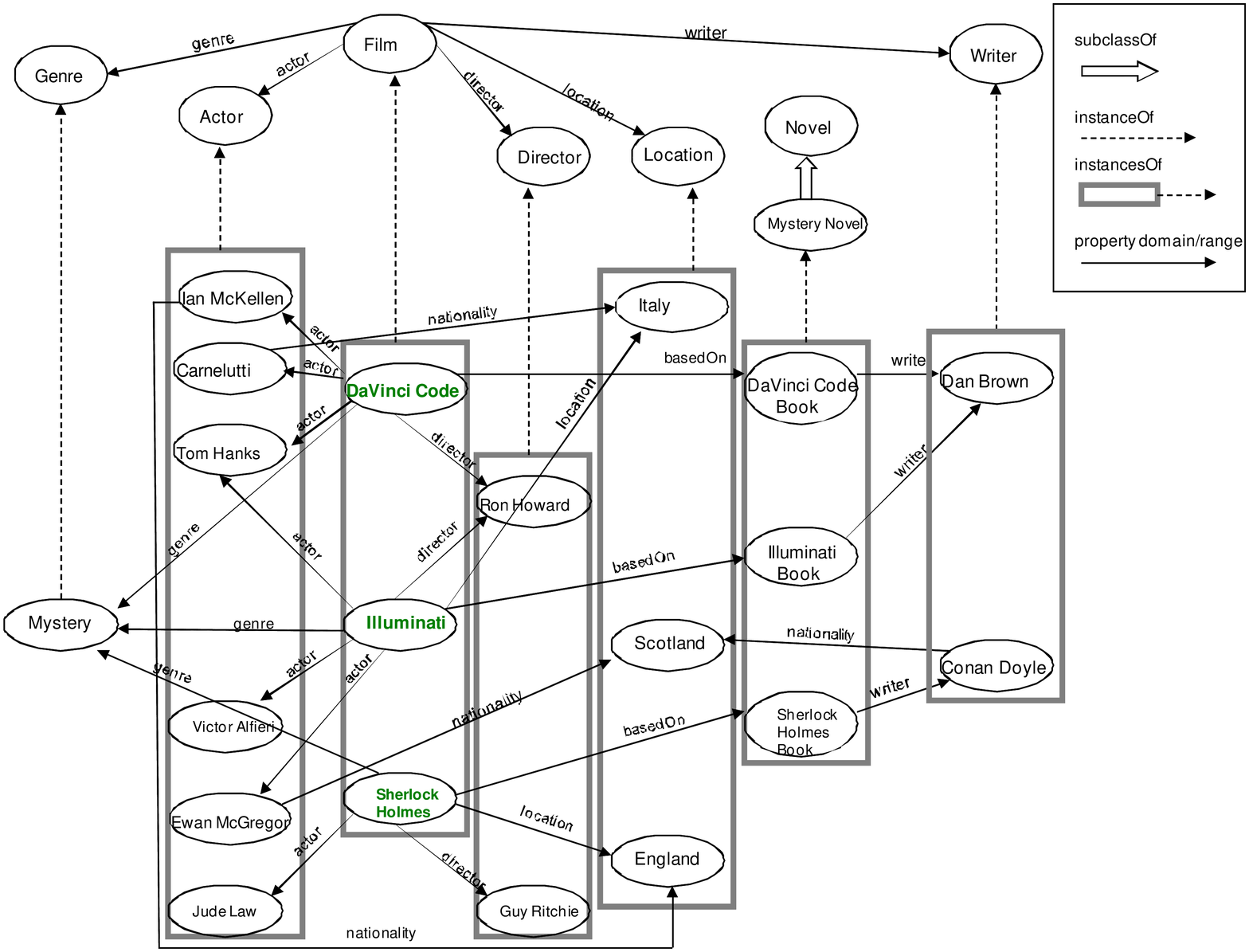}}}
    \caption{The RDF graph $G$ of our running example}\label{fig:fig1}
\end{figure*}

\section{Background (RDF definitions and notations)}
\label{sec:Notations}

An RDF Knowledge Base (KB) is defined as a set of RDF triples,
denoted by  $K$, each having  the form (subject, predicate, object), for short
$(s,p,o)$.
A KB $K$ can also be viewed as a directed labeled graph $G=(N,E)$. The
nodes of the graph are the URIs, the literals and the blank nodes
that appear  in the triples of $K$, while the edges of the graph are
labeled arcs that connect the corresponding nodes.

We shall use as running example the KB that is illustrated
at Figure \ref{fig:fig1}.
For the sake of completeness,
even if the LOD dataset did not have an explicitly defined schema,
we have created one
(for capturing the general case of RDF/S KBs).
Furthermore, we added some extra entities
\footnote{
Specifically
 {\tt DaVinciCode Book},
        {\tt Illuminati Book},
        {\tt Sherlock Holmes Book},
        {\tt Dan Brown},
        and
        {\tt Conan Doyele}.
}
apart from those fetched from LOD.

\comment{
  {\bf
        (TO CHANGE:)
        We should note that we created the RDFS schema
        based on the information
        available from LinkedMDB.
        Apart from the illustrated classes
        at Fig. \ref{fig:fig1}
        we have added the properties
        {\tt location, nationality, basedOn, writer}.
        Moreover,
        we added some information at instance level.
        For instance,
        and with respect to our running example that will be presented
        later, we added the resources
        {\tt DaVinciCode Book},
        {\tt Illuminati Book},
        {\tt Sherlock Holmes Book},
        {\tt Dan Brown},
        and
        {\tt Conan Doyele}.
    }
}

All resources which are instances of a class
are vertically aligned with the class.
Below we introduce some notations which
are necessary for defining the similarity metric.

We shall use $Pr$ to refer to the properties that occur in $K$.
For a given resource $u$ we shall use $ResFrom(u)$ (resp. $ResTo(u)$) to denote
the resources which are pointed to by (resp. point to) resource $u$, i.e.
\bequa
 ResFrom(u) &=& \{~o~|~(u,p,o)\in K, p \in Pr\} \\
 ResTo(u)   &=& \{~o~|~(o,p,u)\in K, p \in Pr\}
\eequa
In our running example we have:\\
{\small $ResFrom($ $Sherlock Holmes)=\{ England,Guy Ritchie,$ $Jude Law, Mystery,Sherlock  Holmes Book\}$.}

We define the classes and the superclasses of a resource $u$ as:
\bequa
    Classes(u)      &=& \{~c~|~(u,type,c)\in K \} \\
    SuperClasses(u) &=& \{~c~|~(u, subClassOf,c) \in K\}
\eequa
For example in Figure \ref{fig:fig1} we have: \\
{\small $Classes(Illuminati Book)=\{Mystery Novel\}$}
while
{\small$ SuperClasses(MysteryNovel)=\{Novel\}$.}
Obviously
if an element $x$ is a class then, $Classes(x) =\emptyset$,
while if $x$ is an instance of a class
then $superClasses(x)$ $=\emptyset$.

Some notations for edges follow.
We define the set of classification
and inheritance links of a resource $u$ and a class $c$ as:
\bequa
        ClassLinks(u)   &=& \{~(u,c)~|~(u, type,c) \in K\} \\
        SupLinks(c)     &=& \{~(c,c')|(c, subClassOf,c') \in K\}
\eequa
The inbound and outbound property links  of a resource $u$ are defined as:
\bequa
    PropsFromLinks(u)    =\{~(u,o)~|(u,p,o)\in K, p \in Pr\} \\
    PropsToLinks(u)      =\{~(o,u)~|(o,p,u)\in K, p \in Pr\}
\eequa

Now we extend the above definitions to take as parameter a set $(S)$ of resources, so we have:
{\small
\bequa
ResFrom(S)          &=& \cup_{u\in S} ResFrom(u) \\
ResTo(S)            &=& \cup_{u\in S} ResTo(u) \\
PropsFromLinks(S)   &=& \cup_{u\in S} PropsFromLinks(u) \\
PropsToLinks(S)     &=& \cup_{u\in S} PropsToLinks(u) \\
Classes(S)          &=& \cup_{u\in S} Classes(u) \\
SuperClasses(S)     &=& \cup_{u\in S} SuperClasses(u) \\
ClassLinks(S)       &=& \cup_{u\in S} ClassLinks(u) \\
SupLinks(S)         &=& \cup_{u\in S} SupLinks(u)
\eequa
}

A {\em path over $G$}, is any sequence of edges of the form:
$(A,P,C) , (C,P',D) ,\cdots, (E, P'',u)$,
where all predicates
($P, P',.. P''$)
are either properties in $Pr$ or
the predicate {\tt type} or
the predicate
{\tt subClassOf}.

We define the {\em distance} between two nodes $A$ and $B$ over $G$,
denoted by $dist_G(A,B)$, as the length of the {\em shortest} path
from $A$ to $B$.
If no path exists
then the distance is assumed to be infinite.

\section{Similarity Function}
\label{sec:Similarity}

In this section, we will introduce and
analyze, step by step, the proposed similarity metric,
over the running example of Fig. \ref{fig:fig1}.
Suppose we want to compute the similarity between
two nodes $A$ and
$B$ of the RDF graph $G$.
At first we define the subgraphs of
$A$ and $B$
of {\em radius} $k$, denoted by:
\bequa
 g_{A(k)} &=&(N_k(A),E_k(A)) \\
 g_{B(k)} &=& (N_k(B),E_k(B))
\eequa
They consist of all nodes and edges
that are visited
if we start from  $A$ and $B$ respectively,
and traverse all links
(properties, type, subclassOf)
for depth up to $k$
where the  value of $k$
is  configured externally
(and it will be discussed later on).

These graphs can be computed in an iterative manner.
For instance, for
defining $g_{A(k)}$
we start
from
$g_{A(0)}=(N_0(A),E_0(A))$ where
$N_0(A)=\{A\}$ and $E_0(A) = \emptyset$.
Subsequently, from \\
$g_{A(i-1)} = (N_{i-1}(A),E_{i-1}(A))$
we can compute \\
$g_{A(i)} = (N_i(A),E_i(A))$
(for all $1\leq i \leq k-1$),
as follows:
\bequa
 N_i(A) &=&     N_{i-1}(A) \cup \\
        & &     ResFrom(N_{i-1}(A)) \cup \\
        & &     Classes(N_{i-1}(A)) \cup \\
        & &     SuperClasses(N_{i-1}(A)) \\
 E_i(A) &=&     E_{i-1}(A) \cup \\
        &  &    PropsFromLinks(N_{i-1}(A)) \cup \\
        &  &    ClassLinks(N_{i-1}(A)) \cup \\
        &  &    SupLinks(N_{i-1}(A))
\eequa

Each step
of the iteration
enriches the current set  of nodes $N_{i-1}(A)$ with the nodes:
\bi
\item which are classes of a node in $N_{i-1}(A)$  (since classes carry important information),
\item the values of the properties that start from the nodes in $N_{i-1}(A)$
      (they are actually attribute values),
\item the superclasses of the nodes in $N_{i-1}(A)$
      (for climbing up the subClassOf hierarchy)
\ei

The iterative expansion allows collecting
values of complex attributes,
as well as
higher level superclasses
(in this way we can detect similarities even
    between very "distant" entities
    which belong to different class hierarchies).

We should stress at this point,
that one could adopt a different policy
regarding how a subgraph expands.
For instance,
one could also expand the graph
using properties
which {\em point to} the current set of nodes
(in that case
$ResTo(N_{i-1}(A))$ would be added to $N_i(A)$
and
$PropsToLinks(N_{i-1}(A))$ to $E_i(A)$
).
The decision is application or ontology specific.
\cite{kiefer2007semantic,kiefer2007fundamentals} have also made the observation
that it is often not enough
to use a single similarity measure
to achieve good results,
therefore a combination of features
needs to be engineered or even learned.
In our case we decided to
take only the forward property direction
since in most cases
a property is more important
for its origin than for its destination.

\begin{figure*}[htbp]
    \centerline{\fbox{\includegraphics[height=8cm]{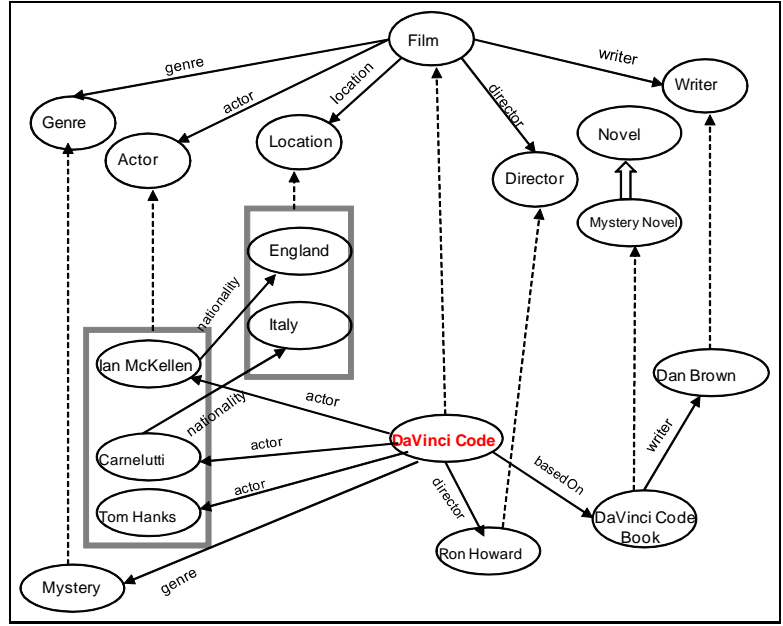}}}
    \caption{ $g_{A(3)}$ where $A$ = {\tt DaVinci Code}}\label{fig:fig2}
\end{figure*}

\begin{figure*}[htbp]
    \centerline{\fbox{\includegraphics[height=9cm]{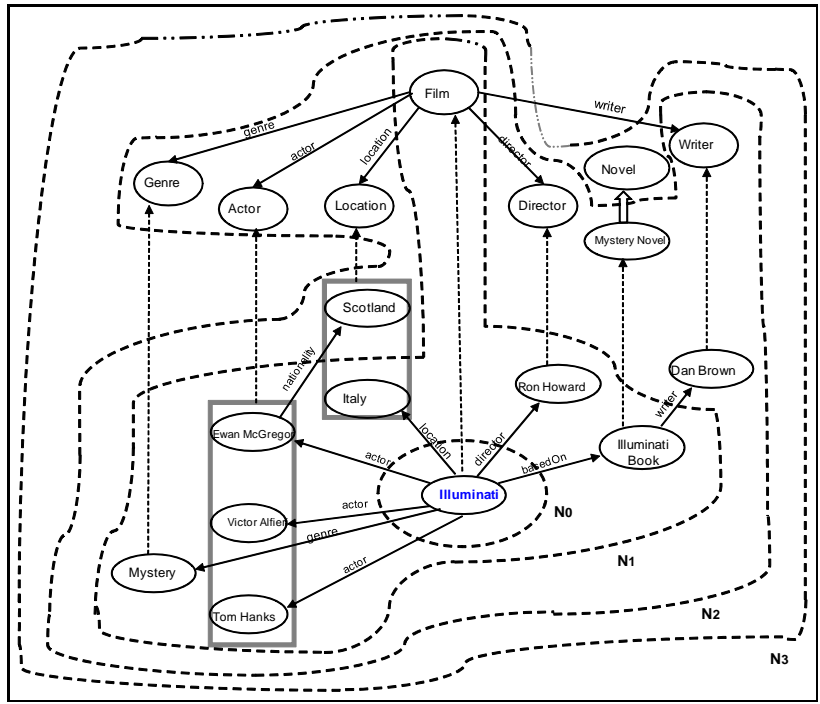}}}
    \caption{$g_{B(3)}$ where $B$ ={\tt Illuminati}}\label{fig:fig3}
\end{figure*}

To better illustrate the construction of the subgraph,
consider the graph $G$  of  Figure \ref{fig:fig1}
and suppose that  $A$ = {\tt DaVinci Code} and $B$ = {\tt Illuminati}.
The subgraphs  $g_{A(3)}$ and $g_{B(3)}$
are shown at Figure \ref{fig:fig2} and Figure \ref{fig:fig3} respectively
(the latter  depicts all subgraphs for $k=0$ to $k=3$).

\begin{figure*}[htb]
    \centerline{\fbox{\includegraphics[height=77mm]{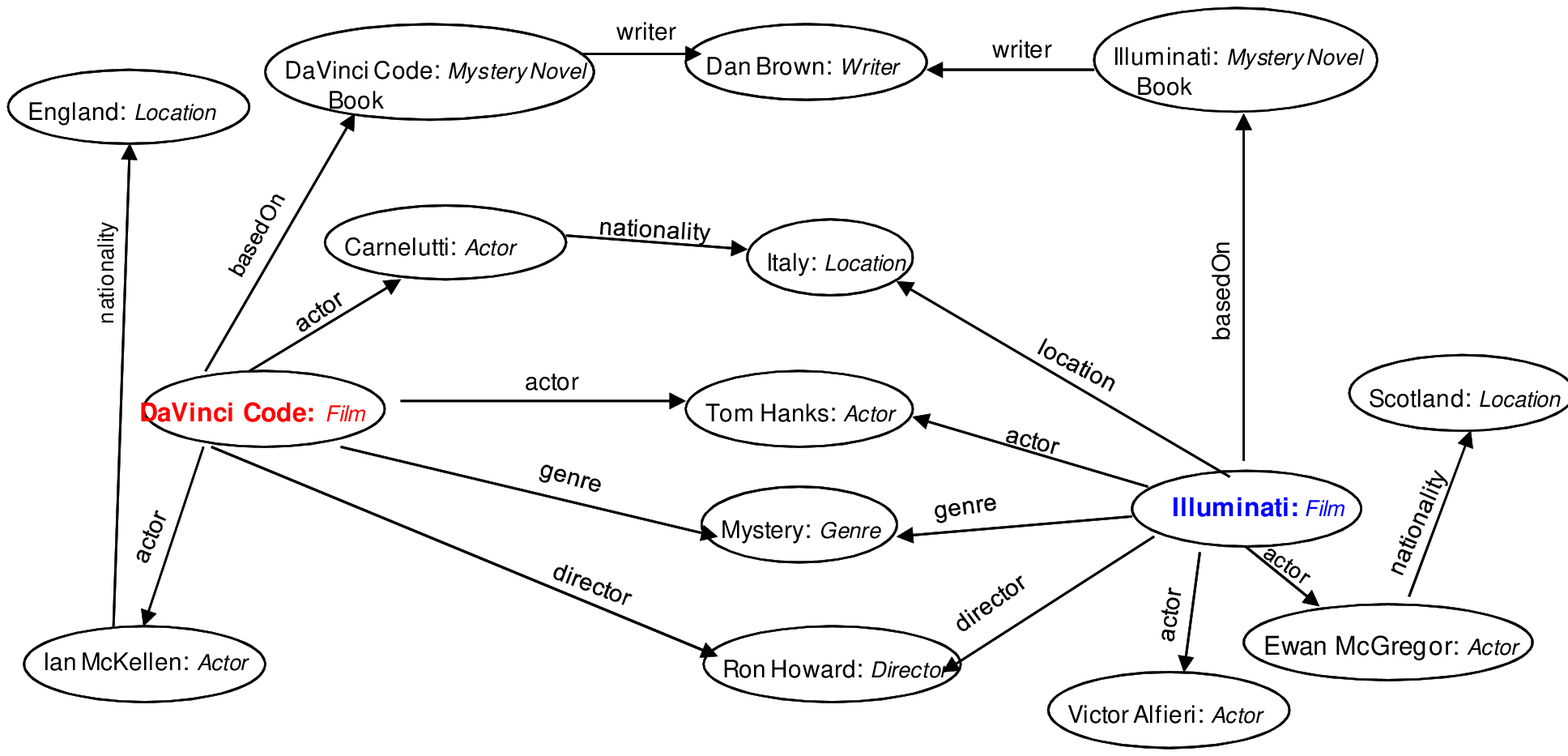}}}
    \caption{Intersection between Illuminati and DaVinci Code Subgraphs}\label{fig:fig4}
\end{figure*}

Table \ref{tbl:table1} shows the distances  $dist_{g_A}(A,u)$ and
$dist_{g_B}(B,u)$
for various $u$ nodes.
%
The nodes for which  both $dist_{g_A}(A,u)$ and $dist_{g_B}(B,u)$
are defined
(i.e. both are different than $\infty$),
actually belong to the intersection of the nodes
of the two subgraphs, while
the rest are nodes that belong only to one of the subgraphs.

\begin{table}[htb]
        \begin{tabular}{|l*{1}|{c}|r|}\hline
            $u$              & $dist_{g_A}(A,u)$ & $dist_{g_B}(B,u)$ \\\hline
            \hline
            Genre                       & 2 & 2 \\\hline
            Actor               & 2 & 2 \\\hline
            Film                  & 1 & 1 \\\hline
            Director                & 2 & 2 \\\hline
            Location                    & 2 & 2 \\\hline
            Novel               & 3 & 3 \\\hline
            Mystery Novel         & 2 & 2 \\\hline
            Writer                      & 2 & 2 \\\hline
            Mystery           & 1 & 1 \\\hline
            Ian McKellen          & 1 & $\infty$ \\\hline
            Carnelutti          & 1 & $\infty$ \\\hline
            Tom Hanks               & 1 & 1 \\\hline
            Victor Alfiery    & $\infty$ & 1 \\\hline
            Ewan McGregor     & $\infty$ & 1 \\\hline
            Ron Howard          & 1 & 1 \\\hline
            Italy                   & 2 & 1 \\\hline
            Scotland                    & $\infty$ & 2 \\\hline
            England           & 2 & $\infty$ \\\hline
            DaVinci Code Book   & 1 & $\infty$ \\\hline
            Illuminati Book   & $\infty$ & 1 \\\hline
            Dan Brown           & 2 & 2 \\\hline
           \end{tabular}
        \caption{Distances from $A$ and from $B$}\label{tbl:table1}
\end{table}

After having constructed
the graphs
$g_A$ and $g_B$ ,
one could compute the similarity between $A$ and $B$
by applying the Jaccard
similarity coefficient \cite{Jain:1988:ACD:46712}
over their node sets, i.e. between $N(A)$ and $N(B)$, as follows:

\begin{equation} \label{eq:type1}
    sim_k(A,B)= \frac{ | N_k(A) \cap  N_k(B) | }  { | N_k(A) \cup  N_k(B) | }
\end{equation}

In our example the  intersection between $N_3(A)$ and $N_3(B)$
is illustrated (vertically aligned) at the center of Figure \ref{fig:fig4}
where
for reasons of space we do not show the schema level intersections.

\comment{
    This function (\ref{eq:type1}) calculates the intersection between all nodes of the
    two subgraphs in depth $\kappa$, initially defined, and is an
    extension of the Jaccard Similarity Coefficient
    \footnote{
        The Jaccard Coefficient which measures the similarity between A and B:
        \begin{equation} \label{eq:type2}
            \frac{|A \cap B|}{|A \cup B|}
        \end{equation}
    }
    , which is widely used in the area of Information Retrieval, with the
    difference that it calculates intersections in depth greater or
    equal than 1$(\kappa>=1)$ as opposed to function (\ref{eq:type2}).
}

Note that by considering the nodes at depth greater than $1$,
we can identify similarities between resources of different types.
If resources of different
types are compared (e.g. a film with an actor),
they will rarely
have the same properties in small depth (e.g. for $k=1$) and
therefore we will not get many (or any) intersecting nodes.

Obviously the similarity value obtained
depends on the value of $k$.
For example,
for $k=1$ we get:
\begin{center}
$sim_1(DaVinci Code, Illuminati) = \frac{4}{15}=0.26$
\end{center}
while for $k=3$  we get
\begin{center}
$sim_3(DaVinci Code, Illuminati) = \frac{13}{21}=0.61$
\end{center}
\comment{
    If we observe the results  we can see that function (\ref{eq:type2}) shows more
    similarity between the two movies, as it compares a wider range of
    nodes.
}

However a shortcoming of this approach, is that a
common node spotted at depth $1$,
is equally weighted as a
common node of a larger distance.
For this reason below we introduce
a different  similarity function
which takes
into account
the values
 $dist_{g_A}(A,u)$ and
$dist_{g_B}(B,u)$.
We should clarify that this extension
does not increase the computational cost
of the similarity function
since these
distances
are computed anyway during the construction
of the subgraphs
$g_{A(k)}$
and
$g_{B(k)}$.

To understand the extension
we shall first
express
function (\ref{eq:type1})
in a different, but equivalent, manner:
\begin{equation} \label{eq:type1variation}
    sim_k(A,B)=
    \frac
        {\sum_{n \in (N_k(A) \cap  N_k(B))} 1}
        {\sum_{n \in (N_k(A) \cup  N_k(A))} 1}
\end{equation}
This form makes evident that each element
in the intersection or union contributes the value of one.
Now we will introduce the new
formula
in which
each
element
in the intersection or union does not contribute the value of one,
but a value
based on its average distance from nodes $A$ and $B$.

Since the closest node is at distance  1
while the most distant is at distance  $k$ (or infinite)
we shall use the expression
$k+1- dist$
for giving to the closest nodes  a contribution  equal to $k$
and to the more  distant nodes a contribution equal to 1.
If a distance equals $\infty$
we consider it
as $k+1$.
In this way the expression $k+1- dist$ yields a zero\footnote{
    This means
    that the cells
    of Table \ref{tbl:table1} that have an
    infinite value ($\infty$)
    are actually considered
    to have the value $k+1$, i.e. 4.
}.

\comment{
    A rising question is what happens
    if a node $u$ does not belong to one of the graphs
    (say to $g_{A(k)}$)
    In that case,
    and for computing the denominator
    of (\ref{eq:SimGood}),
    we consider that its
    distance is equal to $k+1$
    (i.e. $dist_{g_A}(A,u)=k+1$).
    In this way
    only its distance from the other node $B$
    (to whose graph it belongs to)
    is taken into account
}

The proposed similarity function is defined as:
$sim_k(A,B)=$
\begin{eqnarray}\label{eq:SimGood}
        \frac
        {\sum_{n \in (N_k(A) \cap  N_k(B))}
            \frac{(k'-dist_{g_A}(A,n))+(k'- dist_{g_B}(B,n))} {2}}
        {\sum_{n \in (N_k(A) \cup  N_k(B)}
            \frac{(k'- dist_{g_A}(A,n))+(k'- dist_{g_B}(B,n))} {2}}
\end{eqnarray}
where $k' = k+1$.

\comment{
    For each
    intersection between the two subgraphs we have two distances, one
    from $g_A$ and one from $g_B$, and therefore we calculate the
    average of the nodes distances from each subgraph, so that we can
    extract a more fair and balanced result. For the calculation of the
    average we don't simply use each nodes distance, but the subtraction
    of the nodes distance from $k'$ , so that more weight is given
    to the nodes that are more close to the root node and less to the
    ones that are further.
}

If we apply (\ref{eq:SimGood}) to our running example we now get:
{\small
\begin{center}
$sim_3(DaVinci Code, Illuminati) = \frac{29.5}{42}=0.7$
\end{center}
}

In brief, the proposed similarity between two nodes $A$ and $B$ is actually the
Jaccard similarity coefficient evaluated
over the nodes of the
extended neighborhoods of the compared nodes.
Each node of the neighborhoods is
weighted
so that the nodes closer to the compared nodes
get a greater weight than the distant ones.

\subsection{Properties of the Similarity Function}

For any resource $u$,
and for any positive integer $k$
it holds:
$sim_k(u,u)=1$.

It is also clear
that the metric is symmetric
i.e.
$sim_k(a,b)=sim_k(b,a)$.

Although in the examples that we have seen earlier
it happens to hold:
if $m > m'$ then
$sim_m(a,b) \geq sim_{m'}(a,b)$,
in the general case this  does not hold.
The reason is that
for a high $k$
we may have several non intersecting sets of nodes
which increase the denominator
of the similarity function.
\comment{
    kanontas ena paradeigma eidame oti den isxyei h ekshs idiothta ka8os phrame ta ekshs apotelesmata:
    gia k=1 sim=0.66,
    gia k=2 sim=0.44,
}


\begin{figure*}[htb]
    \centerline{\fbox{\includegraphics[height=41mm]{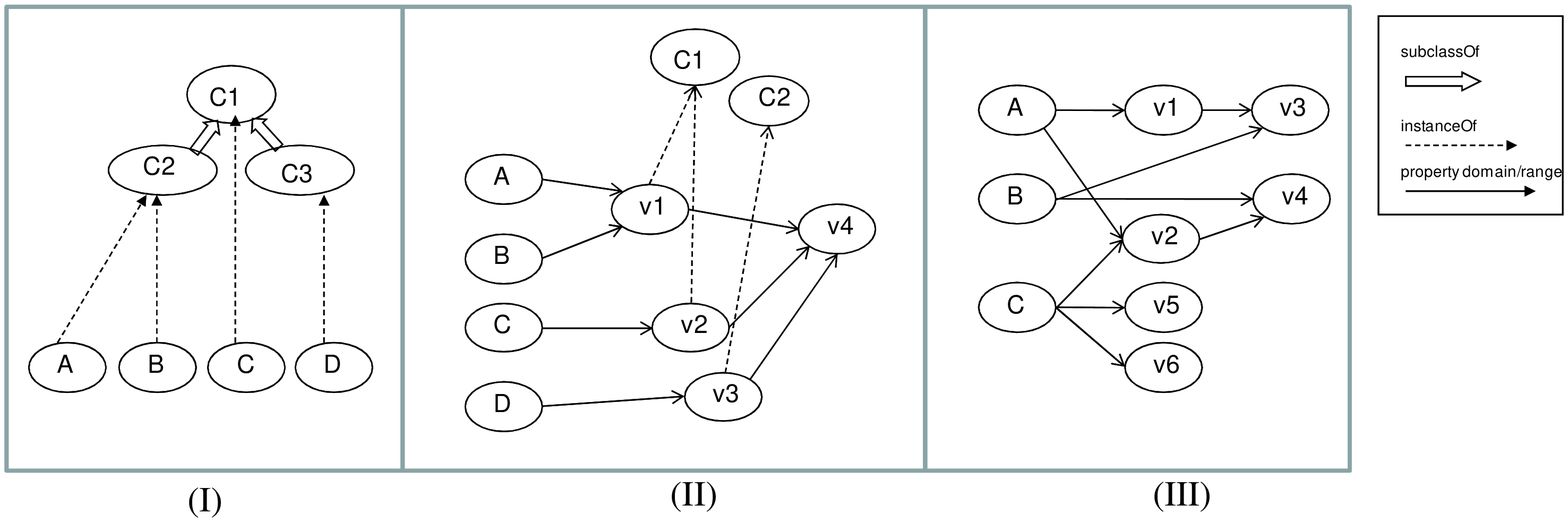}}}
    \caption{Three examples}\label{fig:NewEx}
\end{figure*}

\section{Examples and Analysis}
\label{sec:Evaluation}


\subsection{Behavior}

Table \ref{tbl:table2}
shows the computed similarities
between the films
{\tt DaVinci Code},
{\tt Illuminati} and
{\tt Sherlock Holmes},
for $k = 1, 2, 3$.
We observe that
the most similar movie
with
{\tt DaVinci Code},
is
{\tt Illuminati}
(and not {\tt Sherlock Holmes})
for all values of $k$ from 1 to 3.

\begin{table}[htb]
\bc
\btbl{|l|p{3cm}|p{3cm}|}\hline
$k$  & $sim_k$(DaVinciCode, Illuminati) & $sim_k$(DaVinciCode,
SherlockHolmes) \\\hline
1   & 0.53  & 0.30 \\\hline
2   & 0.67  & 0.54 \\\hline
3   & 0.70  & 0.58 \\\hline
\etbl
\ec
\caption{Similarity for different values of $k$}\label{tbl:table2}
\end{table}

\comment{
    We observe that for smaller $k$
    ($k = 1$) the intersections between the two graphs are very
    few,
    compared to the total number  of nodes and consequently there is
    little similarity between the two films, while as $k$
    increases, despite the fact that the weight of the nodes diminishes,
    the number of intersections increases significantly (taking into
    account the intersections of classes in the RDFS level), therefore
    increasing the similarity between the two nodes.
}

\comment{
    {\bf
    Apla skeftomai oti
    gia na dikaiologhsoume
    to giati exei nohma na exoume $k>1$,
    tha eprepe na exoume ena paradeigma
    to opoio
    to poio similar node se ena A
    na einai allo me $k=1$
    kai allo me $k>1$.
    Alliws kapoios tha mas pei
    giati den xrhsimopoieite k=1.
    Genikh h timh ths omoiothtas den exei synithws shmasia.
    Shmasia exei to ean allazoun
    ta pio similar.

    Isws prepei na ftiaxoume kai ena allo paradeigma
    pou na deixnei tetoia fenomena.
    }
}


Let us now use some examples
to justify the benefits  of $k$ values higher than 1,
and
to better understand the behavior of the similarity function.
Table \ref{tbl:Classes}
shows the computed similarities
between the
nodes
{\tt A},
{\tt B},
{\tt  C} and
{\tt D},
for $k = 1 \ldots 3$,
for the example shown at Figure \ref{fig:NewEx}(I).
We observe that for $k=1$, {\tt B} is the most similar to {\tt A}
since they are under the same class,
while the similarity of {\tt A} with {\tt C} and {\tt D} is zero.
However for  $k=2$
the
similarity of {\tt  A} with {\tt C} and {\tt D} is not zero,
and {\tt C} is more similar than {\tt D}.

\begin{table}[htb]
\bc
\btbl{|l|l|l|l|}\hline
$k$  & $sim_k$(A, B) & $sim_k$(A,C) & $sim_k$(A,D)  \\\hline
1   & 1     & 0     &  0\\\hline
2   & 1     & 0.60  & 0.33   \\\hline
3   & 1     & 0.625  & 0.40\\\hline
\etbl
\ec
\caption{Similarity for different values of $k$ over Fig. \ref{fig:NewEx}(I)}\label{tbl:Classes}
\end{table}

To demonstrate the potential of the similarity function
to exploit commonalities in property paths,
Table \ref{tbl:PropPaths}
shows the computed similarities
between the
nodes
{\tt A},
{\tt B},
{\tt C} and
{\tt D},
for $k = 1 \ldots 3$,
for the example shown at Figure \ref{fig:NewEx}(II).
We observe that
for $k=2$
{\tt A} is more similar to {\tt C}
than to {\tt D}
because
even though they do not have any direct value in common,
{\tt v1} and {\tt v2}
are under the same class {\tt C1},
and {\tt v4} is a common value at depth 2.
Notice that the similarity
between {\tt A} and {\tt D}
is not zero for $k=2$,
due to the value $v4$.

\begin{table}[htb]
\bc
\btbl{|l|l|l|l|}\hline
$k$  & $sim_k$(A, B) & $sim_k$(A,C) & $sim_k$(A,D)  \\\hline
1   & 1      & 0     &  0\\\hline
2   & 1   & 0.50  & 0.25   \\\hline
3   & 1   & 0.57  & 0.28\\\hline
\etbl
\ec
\caption{Similarity for different values of $k$ over Fig.
\ref{fig:NewEx}(II)}\label{tbl:PropPaths}
\end{table}

\comment{
        Q:Mporoume mhpws na vroume kai ena paradeigma tetoio
        wste gia k=1
        to pio similar sto A entity na einai ena entity B
        enw gia ena megalytero k (p.x. gia 2 'h 3)
        to pio similar na mhn einai to B
        alla kapoio allo entity?
}

\comment{
    \begin{figure}[htbp]
        \centerline{\fbox{\includegraphics[height=37mm]{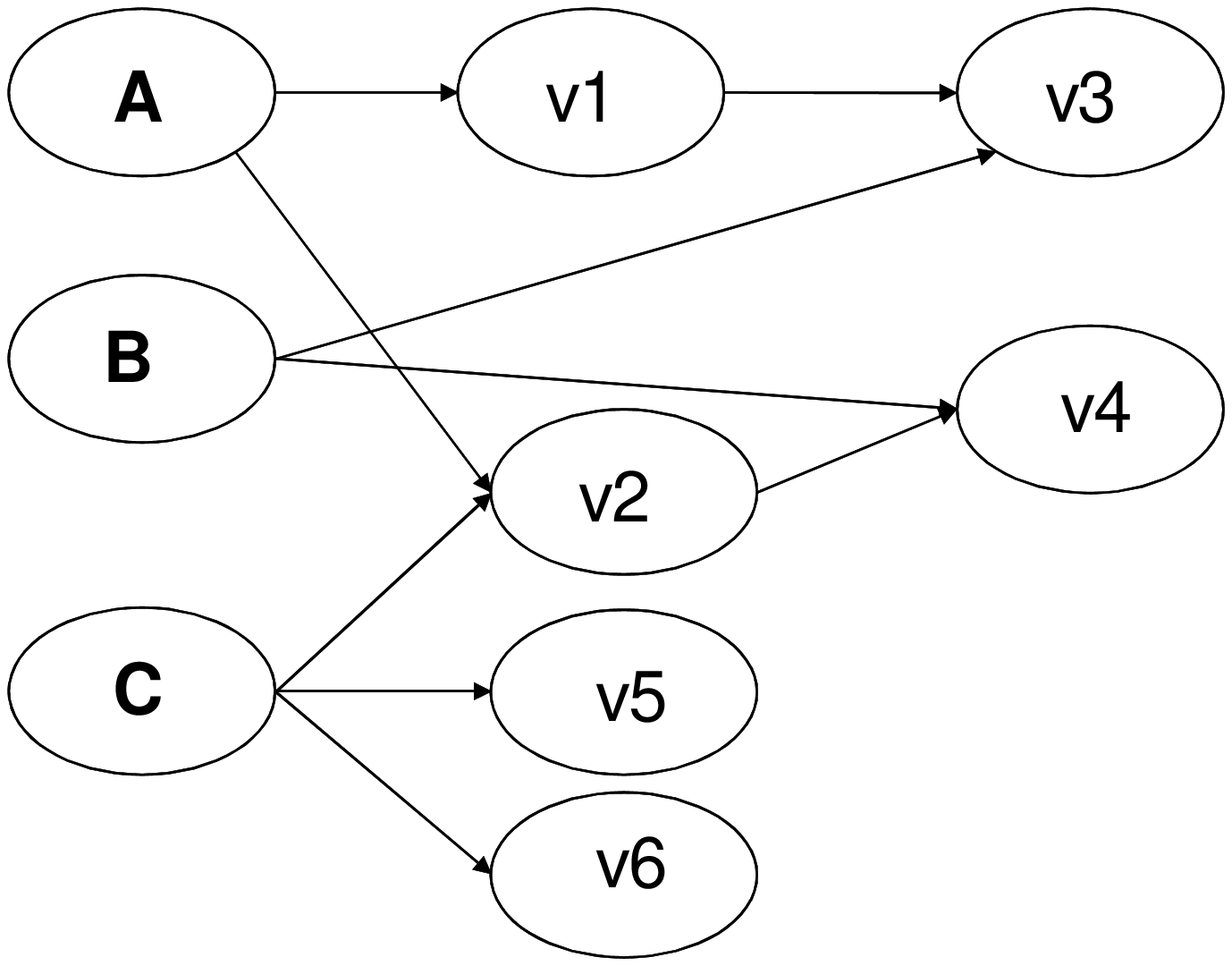}}}
        \caption{Another example}\label{fig:fig6}
    \end{figure}
}

It is also worth noting
that the most similar entity can change as $k$ changes.
For instance, in the example of Figure \ref{fig:NewEx}(III),
as we can see from Table \ref{tbl:table5},
for  $k =1$ the most similar to $A$ is the entity $C$,
while for $k=2$ (and higher) the most similar to $A$ is the entity $B$.

\begin{table}[htb]
\bc
\btbl{|l|l|l|}\hline
$k$  & $sim_k$(A, B) & $sim_k$(A,C)  \\\hline
0   & 0     & 0         \\\hline
1   & 0     & 0.40     \\\hline
2   & 0.60     & 0.46  \\\hline
3   & 0.625     & 0.47\\\hline
\etbl
\ec
\caption{Similarity for different values of $k$ over Figure \ref{fig:NewEx}(III) }\label{tbl:table5}
\end{table}

\comment{
    We know that the films {\tt DaVinci Code} and {\tt Illuminati} are
    quite similar, which confirms our function's results ($sim = 0.7$).
    Another film that is similar (less than {\tt Illuminati}) to {\tt
    DaVinci Code} is {\tt Sherlock Holmes}.
    If we apply our function to
    these two films for $k=3$, as before, we will get the following
    result:
    {\small
    \begin{center}
    $sim_3({ DaVinci Code, Sherlock Holmes}) = \frac{23.5}{39}=0.6,$
    \end{center}
    }
    which confirms the correctness and liability of our function, since
    it finds a relatively good similarity between the two films, but
    less than the one with {\tt Illuminati} (where similarity was
    $0.7$). The reasons why this happens are two: (1) there are fewer
    intersections between the two subgraphs, compared to the total
    number of nodes of the two subgraphs and (2) from all the
    intersections the majority is in greater depth, therefore giving
    less weight to the computation of the similarity function.
}

\subsection{Computational Complexity}

Let $d$ be the average number of edges which are adjacent to a node.
For a node $A$,
the number of nodes
in the  graph $g_{A(k)}$
is at most in  $\cal{O}$$(d^k)$.
This is therefore the cost of $sim_k(\cdot,\cdot)$.

\subsection{On Selecting a value for $k$}

One issue that plays an important role in  the computation of similarity
is the choice of the appropriate $k$.
The choice can be made
by the application designer (or even by the end user at run-time).
By
choosing a greater $k$
more complexity is added to the
computation  of the similarity and
this is the cost to pay
for more accurate results
in the sense that  a wider part of the graph is taken into account.
By choosing a lower $k$ the computational cost  gets decreased,
but the results may not be as accurate as the user would like.

One method for selecting a $k$
is to measure graph features of the RDF/S graph,
e.g. the diameter of the graph.

\comment{
    for example if he knew in advance that most of the
    similarities were in smaller depth he would choose a smaller
    $\kappa$ improving the speed and efficiency of the system.
    As we can
    see, one of the great advantages of the function we suggest is the
    versatility it offers, as it adapts to the needs of each user, in
    contrast to most other approaches where the user must adapt to the
    needs of the function.
}

\subsection{Variations of the Similarity Function}
\label{sec:Variation}

As one may have noticed,
the similarity function
ignores the names of the properties.

The benefit of this choice
is that the function can yield positive similarities
also between objects that use different properties.
For example
consider the triples
{\tt (a, hasFriend, e)} and
{\tt (b, worksFor, e)}.
The similarity function will
return a positive value for $sim_1(a,b)$
although these entities have different properties.
It would be zero
if the property names were taken into account.
However, the shortcoming is
inability to promote matches also at the properties.
For example,
if we had another triple
{\tt (c, hasFriend, e)}
then
we would have
$sim_1(a,c) = sim_1(a,b)$,
although we would prefer
$sim_1(a,c) > sim_1(a,b)$.

If we wanted to take into account the property
names then
we could {\em prefix} the
nodes of the subgraphs
which are reached from properties
by the corresponding property name.
In particular, instead of \\
$ResFrom(u) = \{~o~|~(u,p,o)\in K, p \in Pr\}$, we could define \\
$ResFrom'(u) = \{~p:o~|~(u,p,o)\in K, p \in Pr\}$, \\
where "$p:o$" is treated as one string.
Clearly,
with such a change,
the new similarity function,
denoted by $sim'$,
would yield
$sim'_1(a,c) > sim'_1(a,b)=0$.

One approach to reconcile the two approaches
is to change
the graph expansion step
so that both
$ResFrom(u)$ and  $ResFrom'(u)$
are used
for the definition of the nodes of the subgraphs.
Specifically
$N_i(A)$ can now be defined as:
\bequa
 N_i(A) &=&     N_{i-1}(A) \cup \\
        & &     ResFrom(N_{i-1}(A)) \cup \\
        & &     ResFrom'(N_{i-1}(A)) \cup \\
        & &     Classes(N_{i-1}(A)) \cup \\
        & &     SuperClasses(N_{i-1}(A))
\eequa
In this way we will get
$sim''_1(a,c) > sim''_1(a,b) > 0$.

\subsection{Experimental Results}
\label{sec:Measurements}

\comment{
        \noindent
        {\bf Approach (I):}\\
        {\em
        We download a lot of movies, say 1000,
        and then for a selected subset of them (say 20),
        we compute and show the top-5 most similar
        movies (and/or actors, directors, etc)
        for each of the 20 movies,
        for various similarity functions,
        e.g. for $sim_k$ for $k=1..3$,
        as well as for a COMPETITIVE
        (a good alternative metric - ideally one from the recent literature).
        We present the list in the paper
        and let the reader
        judge by himself whether the
        similar pages make sense
        and which of the two similarity functions seems to behave better.
        }
}

We created a bigger KB for
testing the similarity function,
i.e. for judging whether it returns intuitive results
and for investigating how the value of $k$ affects the results.

\noindent
[{\em Setup of the KB }] \\
Our measurements were based on a KB that we created by
extracting data  from LinkedMDB, through Virtuoso's SPARQL Endpoint,
with explicit queries. More specifically, we selected and downloaded 10 entities,
that were quite relevant to each other.
For  each one of them we
expanded their subgraphs for depth 3,
and with the fetched information
we created a  KB on which
our measurements were conducted.
The entities that were chosen and
their types are shown in
Table \ref{tbl:SelectedEntities}.

The  resulting KB contained:
16 classes,
70 properties,
3326 resources,
4301 property instances,
and 4877 triples in sum.
\comment{
    ISSUE:
    Kanonika tha eprepe na eixe mpei kai to sxhma.
    Xwris auto sth ousia sto subclass expansion
    den egine stis metrhseis mas.
    Vevaia apo oti eida sta figures, eixame mono ena: MysteryNovel subclassOf Novel.
    Apo thn allh ola ta similarities tha htan diaforetika
    afou tha eixame matches sto sxhma.
}

\begin{table}[htb]
\bc
\btbl{|l|l|} \hline
Angels and Demons   & Film \\\hline
The DaVinci Code    & Film \\\hline
That Thing You Do!  & Film \\\hline
Original Sin        & Film \\\hline
Jude                & Film \\\hline
Catch Me If You Can & Film \\\hline
Leonardo DiCaprio   & Actor \\\hline
Tom Hanks           & Actor \\\hline
Phil Alden Robinson & Director \\\hline
Joe Dante           & Director \\\hline
\etbl
\ec
\caption{Selected (seed) entities}\label{tbl:SelectedEntities}
\end{table}

\comment{
Kalhmera sas,
H monh sxetiki synartisi p vrikame einai h size() h opoia epistrefei
ton ari8mo olon ton tripleton toy montelou kai mas epistrefei 4877
triples. Se aytes logika symperilambanontai kai oi tripletes me typeof
alla pera apo aytes den yparxoyn alles sysxetiseis se rdfs epipedo(px
subclassof...). Ara oi tripletes xoris tis klaseis einai 4877-576=4301
,  opou 576 einai oles oi tripletes ths morfhs a typeof b.
}

\comment{ INTERNAL
    gia ta statistika treksame ena programma sto opoio mesa apo 10 resources epelegei 2 tixaia kai vriksei to
    sim metaksu ayton ton duo. Ayti i diadikasia egine  100 fores gia ka8e k. Sti sinixeia vriskei to meso oro
    ton sim pou vrike 100 fores gia ka8e k.
    Paratiroume oti an to sim einai 0, dld den iparxei kaneis komvos o xronos einai iso me 0, enw an to sim einai
    1, dld sigkrinoume to idio resource episis o xronos einai 0.
    Sas episinaptoume mia endiktiki eksodo gai k=3 pou eixe to programa
}


\noindent
[{\em Top-3 Results]}\\
We computed the similarity between every pair
of these 10 entities
for all  $k=1, 2, 3$.
Table \ref{tbl:CompResults} shows the top-3 most similar entities for each entity.

\begin{figure*}
\bc
{\scriptsize
\btbl{|p{29mm}||p{31mm}|p{31mm}|p{31mm}|}\hline
Entity & \multicolumn{3}{|c|}{Top-3 more similar entities} \\\hline\hline
     & $sim_1$ & $sim_2$ & $sim_3$  \\\hline
The Da Vinci Code
   &   $\langle$ Angels and Demons, \newline That Thing You Do!,
\newline Catch Me if You Can $\rangle$
   &   $\langle$ Angels and Demons, \newline That Thing You Do!,
\newline Catch Me if You Can $\rangle$
   &   $\langle$ Angels and Demons, \newline Catch Me if You Can,
\newline That Thing You Do! $\rangle$
   \\\hline
Angels and Demons
   &   $\langle$ The Da Vinci Code, \newline That Thing You Do!,
\newline Catch Me if You Can $\rangle$
   &   $\langle$ Tom Hanks, \newline The Da Vinci Code, \newline That
Thing You Do! $\rangle$
   &   $\langle$ Tom Hanks, \newline The Da Vinci Code, \newline That
Thing You Do! $\rangle$
   \\\hline
Tom Hanks
   &   $\langle$ Leonardo DiCaprio, \newline Phil Alden Robinson,
\newline Joe Dante $\rangle$
   &   $\langle$ Angels and Demons, \newline Leonardo DiCaprio,
\newline That Thing You Do! $\rangle$
   &   $\langle$ Angels and Demons, \newline Leonardo DiCaprio,
\newline That Thing You Do! $\rangle$
   \\\hline
That Thing You Do!
   &   $\langle$ Catch Me if You Can, \newline Angels and Demons,
\newline The Da Vinci Code $\rangle$
   &   $\langle$ Catch Me if You Can, \newline Tom Hanks, \newline
Angels and Demons $\rangle$
   &   $\langle$ Phil Alden Robinson, \newline Tom Hanks, \newline
Angels and Demons $\rangle$
   \\\hline
Original Sin
   &   $\langle$ Jude, \newline Angels and Demons, \newline That Thing
You Do! $\rangle$
   &   $\langle$ Jude, \newline Angels and Demons, \newline That Thing
You Do! $\rangle$
   &   $\langle$ Jude, \newline Angels and Demons, \newline The Da
Vinci Code $\rangle$
   \\\hline
Jude
   &   $\langle$ That Thing You Do!, \newline Angels and Demons,
\newline Original Sin $\rangle$
   &   $\langle$ Angels and Demons, \newline Original Sin, \newline
That Thing You Do! $\rangle$
   &   $\langle$ Phil Alden Robinson, \newline Angels and Demons,
\newline Original Sin $\rangle$
   \\\hline
Catch Me if You Can
   &   $\langle$ That Thing You Do!, \newline The Da Vinci Code,
\newline Angels and Demons $\rangle$
   &   $\langle$ That Thing You Do!, \newline The Da Vinci Code,
\newline Angels and Demons $\rangle$
   &   $\langle$ Joe Dante, \newline The Da Vinci Code, \newline That
Thing You Do! $\rangle$
   \\\hline
Leonardo DiCaprio
   &   $\langle$ Tom Hanks, \newline Phil Alden Robinson, \newline Joe
Dante $\rangle$
   &   $\langle$ Tom Hanks, \newline Catch Me if You Can, \newline
Angels and Demons $\rangle$
   &   $\langle$ Tom Hanks, \newline Angels and Demons, \newline Catch
Me if You Can $\rangle$
   \\\hline
Phil Alden Robinson
   &   $\langle$ Joe Dante, \newline Tom Hanks, \newline Leonardo
DiCaprio $\rangle$
   &   $\langle$ That Thing You Do!, \newline Catch Me if You Can,
\newline Angels and Demons $\rangle$
   &   $\langle$ That Thing You Do!, \newline Catch Me if You Can,
\newline Jude $\rangle$
   \\\hline
Joe Dante
   &   $\langle$ Phil Alden Robinson, \newline Tom Hanks, \newline
Leonardo DiCaprio $\rangle$
   &   $\langle$ Catch Me if You Can, \newline Phil Alden Robinson,
\newline That Thing You Do! $\rangle$
   &   $\langle$ Catch Me if You Can, \newline The Da Vinci Code,
\newline Phil Alden Robinson $\rangle$
   \\\hline
\etbl
}
\ec
\caption{Comparative Results for $sim$}\label{tbl:CompResults}
\end{figure*}

\begin{figure*}
\bc
{\scriptsize
\btbl{|p{29mm}||p{31mm}|p{31mm}|p{31mm}|}\hline
Entity & \multicolumn{3}{|c|}{Top-3 more similar entities} \\\hline\hline
   & $sim_1$ & $sim_2$ & $sim_3$  \\\hline

The Da Vinci Code
 &   $\langle$ Angels and Demons, \newline That Thing You Do!,
\newline Catch Me if You Can $\rangle$
 &   $\langle$ Angels and Demons, \newline That Thing You Do!,
\newline Catch Me if You Can $\rangle$
 &   $\langle$ Angels and Demons, \newline That Thing You Do!,
\newline Catch Me if You Can $\rangle$
 \\\hline
Angels and Demons
 &   $\langle$ The Da Vinci Code, \newline That Thing You Do!,
\newline Catch Me if You Can $\rangle$
 &   $\langle$ Tom Hanks, \newline The Da Vinci Code, \newline That
Thing You Do! $\rangle$
 &   $\langle$ Tom Hanks, \newline The Da Vinci Code, \newline That
Thing You Do! $\rangle$
 \\\hline
Tom Hanks
 &   $\langle$ Leonardo DiCaprio, \newline Phil Alden Robinson,
\newline Joe Dante $\rangle$
 &   $\langle$ Angels and Demons, \newline Leonardo DiCaprio,
\newline That Thing You Do! $\rangle$
 &   $\langle$ Angels and Demons, \newline Leonardo DiCaprio,
\newline That Thing You Do! $\rangle$
 \\\hline
That Thing You Do!
 &   $\langle$ Catch Me if You Can, \newline The Da Vinci Code,
\newline Angels and Demons $\rangle$
 &   $\langle$ Catch Me if You Can, \newline Tom Hanks, \newline
Angels and Demons $\rangle$
 &   $\langle$ Phil Alden Robinson, \newline Tom Hanks, \newline
Angels and Demons $\rangle$
 \\\hline
Original Sin
 &   $\langle$ Jude, \newline Angels and Demons, \newline That Thing
You Do! $\rangle$
 &   $\langle$ Jude, \newline Angels and Demons, \newline That Thing
You Do! $\rangle$
 &   $\langle$ Jude, \newline Angels and Demons, \newline The Da
Vinci Code $\rangle$
 \\\hline
Jude
 &   $\langle$ That Thing You Do!, \newline Angels and Demons,
\newline Original Sin $\rangle$
 &   $\langle$ That Thing You Do!, \newline Angels and Demons, \newline
Original Sin $\rangle$
 &   $\langle$ Phil Alden Robinson, \newline That Thing You Do!,
\newline Angels and Demons $\rangle$
 \\\hline
Catch Me if You Can
 &   $\langle$ That Thing You Do!, \newline The Da Vinci Code,
\newline Angels and Demons $\rangle$
 &   $\langle$ That Thing You Do!, \newline The Da Vinci Code,
\newline Angels and Demons $\rangle$
 &   $\langle$ Joe Dante, \newline The Da Vinci Code, \newline That
Thing You Do! $\rangle$
 \\\hline
Leonardo DiCaprio
 &   $\langle$ Tom Hanks, \newline Phil Alden Robinson, \newline Joe
Dante $\rangle$
 &   $\langle$ Tom Hanks, \newline Catch Me if You Can, \newline
Angels and Demons $\rangle$
 &   $\langle$ Tom Hanks, \newline Angels and Demons, \newline Catch
Me if You Can $\rangle$
 \\\hline
Phil Alden Robinson
 &   $\langle$ Joe Dante, \newline Tom Hanks, \newline Leonardo
DiCaprio $\rangle$
 &   $\langle$ That Thing You Do!, \newline Catch Me if You Can,
\newline Angels and Demons $\rangle$
 &   $\langle$ That Thing You Do!, \newline Catch Me if You Can,
\newline Jude $\rangle$
 \\\hline
Joe Dante
 &   $\langle$ Phil Alden Robinson, \newline Tom Hanks, \newline
Leonardo DiCaprio $\rangle$
 &   $\langle$ Catch Me if You Can, \newline Phil Alden Robinson,
\newline That Thing You Do! $\rangle$
 &   $\langle$ Catch Me if You Can, \newline The Da Vinci Code,
\newline Phil Alden Robinson $\rangle$
 \\\hline
\etbl
}
\ec
\caption{Comparative Results for $sim''$}\label{tbl:CompResultsVariation}
\end{figure*}

We can observe that for some entities,
the 3 most similar entities
change when $k$ changes.
For example,
the 3 most similar entities
for {\tt Tom Hanks}
and $k=1$, are: \\
$\langle$
{\tt Leonardo DiCaprio}, \\
{\tt Phil Alden Robinson}, \\
{\tt Joe Dante }$\rangle$ \\
while for $k=2, 3$ they are: \\
$\langle$
{\tt Angels and Demons}, \\
{\tt Leonardo DiCaprio},\\
{\tt That Thing You Do!}$\rangle$.

We also observed that for $k=1$
for some entities we could not get any similar entity.
Therefore higher values of $k$ are beneficial.

\godown

\noindent
[{\em Comparison with  $sim''$ }]

At Section \ref{sec:Variation}
we described a variation of the similarity function,
denoted by $sim''$.
Table \ref{tbl:CompResultsVariation}
shows again the top-3 most similar entities
(as in Table \ref{tbl:CompResults})
when using $sim''$.
We observe that
the results are quite similar to those of Table \ref{tbl:CompResults},
in most times only the relative ordering of the three more
similar entities differs.

\godown

\noindent
[{\em Times]}\\
The average time to compute $sim_k()$ between two randomly selected
resources,
for $k=2$ equals 3 milliseconds,
while
for $k=3$
equals 32 milliseconds.
All experiments were carried out in a computer with processor Intel(R) Core(TM)2 Duo @2.40GHz,
2 GB Ram, running Microsoft Windows 7 Ultimate.

\section{Implementation Approaches}
\label{sec:Implementation}

Here we discuss implementation issues.

\ \\
\noindent
{\em [The Straightforward approach]}\\
One could attempt
to compute
the similar entities
at run-time
during the construction of the page at hand.
However,
that would not be efficient
in the sense that
a lot of information
would have to be fetched
and processed.
In particular,
to compute the similar entities
for an entity $A$
we should compute
the values $sim_k(A,x)$
for all possible resources $x$.
The cost could be reduced by
limiting the set of values that $x$ may take.
Specifically, we can
first specify the classes of the possible
similar entries,
in our case
the classes of
{\em actors, directors, editors, movies, writers}
(as we described at Section \ref{sec:AC}),
and then download all information available only for these resources.
In any case
that would be unacceptably slow and inefficient for large KBs.

\ \\
\noindent
{\em [The Single Repository (and Preprocessing) approach]}\\
An alternative approach
is to download and  process the entire KB
(e.g. as we did in the previous section).
Since
for each entity
we need to
show only the $L$  (e.g. $L$=5)
most similar entities,
we can compute offline
the
$L$ most similar entities
for each entity of the classes of interest,
and then
store these
$L$ resources (e.g. in main memory)
for immediate use at run time.
Recall that current WSE (Web Search Engines)
also compute off-line and store
for each page the 20 most similar pages.
This preprocessing
can be done offline,
before the deployment of the application,
and it can be periodically redone
as new information
becomes available at LOD.

\ \\
\noindent
{\em [A Similarity-Reversal  approach]}\\
An alternative and more challenging implementation approach
is sketched below.
One could attempt
to "reverse" the similarity function,
i.e.
try  traversing
the graph around $A$
and collect
those entities
which have high chances
to be in the top-$L$ most similar entities,
and compute the similarities only for them.
Such an approach
does not require any preprocesssing
and could be feasible
at run time.
Its feasibility also
depends on how exactly the similarity function
is defined.
Below we will
elaborate on such an approach.
The presented approach
can be applied
to our similarity metric,
as well as
to other  similarity metrics
whose computation
requires analyzing
the subgraphs
of the compared entities.
The ultimate objective
is to devise efficient top-$k$ algorithms
(in the spirit of  \cite{fagin99combining,375567}),
appropriate for graph-based similarity measures.
\comment{
    Q:  poso shmantiko einai auto,
    kai giati einai kalytero apo to offiline computation?
}
Nevertheless, such a  method
cannot be  faster than the
preprocessing
method.
On the other hand,
the benefit of adopting such a method
is that it does not require having access
(or ability to store)
the entire KB.
We should note that \cite{Hartig09}
also proposes to query the Web of Linked Data
by traversing RDF links during run-time
since due to the openness of the LOD space
it may not be possible to know in advance
all data sources that might be relevant
for query answering.
We should stress at this point
that our problem is more difficult
since we do not want to evaluate
a single SPARQL query
but to find the most similar entities
and this in general requires
the evaluation of several queries.

\subsection{On Reversing the Similarity Function}
\label{sec:Reversing}

Consider an entity $A$
and suppose
that we want to compute the more similar entities to $A$.
This requires computing
the subgraphs of $A$ as well as
the subgraphs of the other entities of the KB.
Below we will study this problem by considering
one kind of graph expansion at a time.

\godown
\noindent
$\bullet$ $ResFrom(\cdot)$-graph expansion. \\
Suppose the graph expansion is defined only by $ResFrom(\cdot)$.
It is not hard to see that for each
$x \in  ResTo(ResFrom(A))$
it holds: \\
$ResFrom(A) \cap ResFrom(x) \neq \emptyset$. \\
Let $X_{rf}(A) = ResTo(ResFrom(A))$.
Moreover
if $x' \not\in  X_{rf}(A)$,
then
$ResFrom(A) \cap ResFrom(x') =  \emptyset$.
This means that the nominator of the similarity function is certainly greater than zero
only for these entities.

\godown
\noindent
$\bullet$ $Classes(\cdot)$-graph expansion. \\
For this expansion method,
it is not hard to see that for each
$x \in X_{cl}(A) =
Instances(Classes(A)))$
it holds
$Classes(A) \cap Classes(x) \neq \emptyset$.

\godown
\noindent
$\bullet$ $SupClasses(\cdot)$-graph expansion. \\
Analogously, for each \\
$x \in X_{sp}(A) = SubClasses(SuperClasses(A)))$
it holds \\
$SuperClasses(A) \cap SuperClasses(x) \neq \emptyset$.

\godown


It follows from the above
that all elements of
$X_{\cup}(A) = X_{rf}(A) \cup X_{cl}(A) \cup X_{sp}(A)$,
and {\em only these elements},
have certainly non zero similarity.

Let now discuss the case where $k>1$.
In general a value of $k$ greater than one specifies a  set of expansion {\em paths}.
We can follow these expansion paths to get the nodes
of subgraph for $A$,
and then
"reverse" the expansion paths
and apply
them to the ending nodes
of the graph of $A$.
This should be done with care,
since although a path can have length 3 (i.e. $k=3$),
an ending node of the subgraph could be
the result of an expansion of shorter length (e.g. of one),
implying that reversed paths should be shorter too.

The application of these reversed paths,
can give us the candidate entities.
This is actually what we have described
above for the case where $k=1$.
Below we describe
in detail
this process for any value of $k$.

Consider
the set of strings
$Directions$ =
\{
ResFrom, ResTo, Classes, Instances, SubClasses, SuperClasses\}.
A graph expansion step over RDF/S can be specified
by a subset of this set.
For instance,
the graph expansion  used
by the proposed similarity metric
is specified by the
set \{ResFrom, Classes, SubClasses\}.
We can define the "reverse" of a direction as:
\bequa
Rev(ResFrom) &=& ResTo \\
Rev(Classes) &=& Instances \\
Rev(SubClasses) &=& SuperClasses
\eequa
For a subset  $S \subseteq Directions$,
we define
$Rev(S) = \cup_{s \in S} Rev(s)$.

The Algorithm $getCandidateSimilar$ (shown at Fig. \ref{alg:Rev})
takes as input
a node $A$,
the value $k$,
and a $policy$
being a subset  of $Directions$.
It returns
those objects
which have high chances
to be very similar to $A$
(actually those whose similarity with $A$ is certainly positive)
assuming $sim_k$ over subgraphs defined
using the directions in $policy$.

\begin{figure}[htbp]
\btab
{\bf Algorithm} $getCandidateSimilar$\\
{\em Input}: $A$, $k$, $policy$ \\
{\em Output}: A set of resources\\
(1) $R = \emptyset$;\\
(2)  compute $g_k(A)= (N_k(A),E_k(A))$ w.r.t. $policy$\\
(3) For each $n \in N_k(A)$ \\
(4) \hspace{8mm} let $d = dist(n,A)$ \\
(5) \hspace{8mm} $ R = R \cup traverse(Rev(policy)), n, d)$\\
(6) End for\\
(7) return $R$;
\etab
\vspace*{-5mm}
\caption{Alg. for getting the resources  which have "similar" subgraphs to $A$ using $sim_k$}
\label{alg:Rev}
\end{figure}

At line (2) the algorithm computes the subgraph of $A$
according to the directions set in $policy$.
The distance at line (4) has been computed during line (2).
The invocation
$traverse(dirs, n, d)$
starts from $n$
and follows
the links that correspond
to the argument $dirs$,
for up to distance $d$,
and returns the encountered nodes.
To make it more clear
the set of nodes $N_k(A)$ (at line 2)
can be computed by
$N_k(A) = traverse(policy, A, k)$.
Regarding the correctness of the algorithm,
as explained earlier,
only the elements in the returned $R$
can have non zero similarity to $A$.
After having run the algorithm,
the next step is to compute
$sim_k(A,r)$ for each $r \in R$
and return
the more similar elements.
Specifically, for each $r \in R$
we should get all information returned by
$traverse(policy,r,k)$.
With these information we can compute
$sim_k(A,r)$.
%
This can be done either by code or with queries.
For instance,
$sim_1(A,B)$,
assuming that the subgraphs of $A$ and $B$ are defined
only by $Classes(\cdot)$,
can be computed with a query of the form\footnote{
    To be more precise the
    division has to be casted using XSD data type.
}:

{\footnotesize
\begin{verbatim}
SELECT
   (count(distinct ?class1) as ?intersCard)/
   (count(distinct ?class2) as ?unionCard)
   as ?res WHERE {
    {
      A rdf:type ?class1.
      B rdf:type ?class1.
    } UNION{
        { A rdf:type ?class2. }
        UNION
        { B rdf:type ?class2. }
    }
}
\end{verbatim}
}
\comment{=======
            If subgraphs were defined only $resFrom$,
            then we would have:

            {\small
            \begin{verbatim}
            SELECT (1000* (count(distinct ?res1) as ?tomi)
                    / (count(distinct ?res2) as ?enosi) )
                    as ?res WHERE {
            {
                A ?p1 ?res1.
                B ?p2 ?res1.
                FILTER ( ?p1 != rdf:type && ?p2 != rdf:type)
            }UNION{
                {
                    A ?p3 ?res2.
                    FILTER ( ?p3 != rdf:type)
                }UNION{
                    B ?p4 ?res2.
                    FILTER ( ?p4 != rdf:type)
                }
            }
            }
            \end{verbatim}
            }

=========}

The above query can be extended to capture also the rest
graph expansion steps.
However the  case where $k>1$
requires the formulation of much more
complex queries.
It is easier to do the required
computation with a programming language
than with a query language.

\ \\

We have just seen
how we can collect
only those elements
with positive similarity to $A$,
by
first getting the subgraph of $A$,
and then reversing the expansion paths
that defined the subgraph of $A$.

\godown

\noindent
[{\em Top-$L$ Algorithm}] \\
The above algorithm can be extended
to  become a top-$L$ algorithm,
in case we are interested in finding only the $L$ more similar entities.
Let's start from the case where $k=1$
and suppose that the cardinality of the set
$X_{\cup}(A)$
is high.
Since we are interested in finding the $L$ most similar to $A$ entities,
we can adopt a different, more efficient, evaluation approach,
specifically
we can avoid collecting all elements
that will be fetched at line (5)
of the algorithm $getCandidateSimilar$.
The idea is to collect at first
those elements in
$X(A)_{\cap} = X_{p}(A) \cap X_{cl}(A) \cap X_{sp}(A)$.
Clearly, the elements in $X(A)_{\cap}$
will have a positive summand for each
part of the similarity function,
and thus have high probability
to contain the $L$ most similar entities.
If they are more than
the desired number of objects $L$,
i.e. if $|X_{\cap}(A)|\geq L$,
then we can rank them and
present the $L$ most similar entities.
The benefit of this method,
in comparison to
collecting
the elements
of the entire
$X_{\cup}(A)$
(i.e. line (5)),
is that the elements of
$X_{\cap}(A)$ apart from being less,
they can be fetched efficiently,
specifically
with one query.

For instance,
the set
$X_{rf}(A)$ can be computed by the following SPARQL query:
{\small
\begin{verbatim}
SELECT  ?y
WHERE {  A  ?p1 ?x.
        ?y  ?p2 ?x.
        FILTER ( ?p1 != rdf:type &&
                 ?p2 != rdf:type)  }
\end{verbatim}
}

\comment{===OLD====
    \begin{verbatim}
    SELECT ?y
    WHERE{  A ?p ?x.
        ?y ?p ?x.}
    \end{verbatim}
========}

Note that
if we wanted to use
$ResFrom'$
instead of
$ResFrom$,
then we would have to use the query:
{\small
\begin{verbatim}
SELECT ?y
WHERE { A ?p ?x.
       ?y ?p ?x.
       FILTER ( ?p != rdf:type ) }
\end{verbatim}
}

The set
$X_{cl}(A)$ can be computed by the following SPARQL query:
{\small
\begin{verbatim}
SELECT ?y
WHERE{  A  rdf:type ?x.
        ?y rdf:type ?x.}
\end{verbatim}
}
The set
$X_{sp}(A)$ can be computed by the following SPARQL query:
{\small
\begin{verbatim}
SELECT ?y
WHERE{  A  rdfs:subClassOf ?x.
        ?y rdfs:subClassOf ?x. }
\end{verbatim}
}
Now $X_{rf}(A)\cap X_{cl}(A) \cap X_{sp}(A) $ can be computed by the following SPARQL query:
{\small
\begin{verbatim}
SELECT ?y
WHERE{  A ?p1 ?x.
       ?y ?p2 ?x.

        A  rdf:type ?z.
        ?y rdf:type ?z.

        A  rdfs:subClassOf ?w.
        ?y rdfs:subClassOf ?w.

    FILTER ( ?p1 != rdf:type &&
             ?p2 != rdf:type) }
\end{verbatim}
}
Note that the above query can give a non empty result
only if $A$ is at class level
and thus can have superclasses.

If however
the fetched elements
are less than $L$,
i.e. if $|X_{\cap}(A)|<L$,
then we
have to fetch more elements.
We can start collecting those elements
that belong in intersections of two of the above sets,
i.e.
the elements
in
$X_{p}(A) \cap X_{cl}(A)$,
$X_{p}(A) \cap X_{sp}(A)$, and
$X_{cl}(A) \cap X_{sp}(A)$.

For example,
$X_{rf}(A)\cap X_{cl}(A)$ can be computed by the following SPARQL query:
{\small
\begin{verbatim}
SELECT ?y
WHERE{  A ?p1 ?x.
       ?y ?p2 ?x.

        A  rdf:type ?z.
       ?y  rdf:type ?z.

FILTER ( ?p1 != rdf:type &&
         ?p2 != rdf:type)
     }
\end{verbatim}
}
If again
the fetched elements
are less than $L$,
then we can
collect
those in
$X_{rf}(A) \cup X_{cl}(A) \cup X_{sp}(A)$,
i.e. run the original line (5).
These elements can be fetched using the following
query
{\small
\begin{verbatim}
SELECT ?y
WHERE {
        A ?p1 ?x.
       ?y ?p2 ?x.
        FILTER ( ?p1 != rdf:type &&
                 ?p2 != rdf:type)
      }
      UNION {
        A  type ?z.
       ?y  type ?z.
      }
      UNION {
        A  subClassOf ?w.
       ?y  subClassOf ?w.
      }
\end{verbatim}
}

Essentially
the main idea is the following.
If the subgraph is defined by a set of directions $dirs$,
then instead of reversing each one direction in isolation
and getting the union,
try reversing all directions at once.
Then all directions except one, and so on.
In other words,
it is like
starting from the top node of the Hasse diagram
of the powerset of $dirs$
$(\cal{P}$$(dirs), \subseteq)$
and then descend level wise.
E.g.:

\begin{verbatim}
       {P,C S}         :level 1
       /  |  \
  {P,C} {P,S}{C,S}     :level 2
    | \ / | \/ |
    | / \ | /\ |
   {P}   {C}  {S}      :level 3
\end{verbatim}




\begin{figure*}
{\small
\btbl{|l|r|r|r|r|}\hline
A &
\multicolumn{4}{|c|}{$|Answer(q)|$} \\\hline
 &
$X_{rf \cup cl}(A)$ &
$X_{rf}(A)$ &
$X_{cl}(A)$ &
$X_{rf}(A)\cap X_{cl}(A)$
\\\hline
DaVinciCode
    & 82
    & 76
    & 81
    & 75      \\\hline
Tom Hanks
    & 185
    & 5
    & 182
    & 2      \\\hline
The Thing You Do!
    & 82
    & 75
    & 81
    & 74     \\\hline
\end{tabular}
\caption{Measurements over the local KB}\label{tbl:Reversals}
}
\end{figure*}

\begin{figure*}
{\small
\btbl{|l|r|r|r|r|}\hline
A &
\multicolumn{4}{|c|}{$|Answer(q)|$} \\\hline
 &
$X_{rf \cup cl}(A)$ &
$X_{rf}(A)$ &
$X_{cl}(A)$ &
$X_{rf}(A)\cap X_{cl}(A)$
\\\hline
Americano
    & 1,679,605
    & 32,318
    & 1,679,318
    & 32,031       \\\hline
DaVinciCode
    & 1,683,729
    & 246,918
    & 1,668,503
    & 231,692       \\\hline
Illuminati
    & 1,676,081
    & 98,032
    & 1,668,503
    & 90,454       \\\hline
Tom Hanks
    & 2,218,574
    & 862,458
    & 2,183,320
    & 827,204      \\\hline
\end{tabular}
\caption{Measurements over DBPEDIA}\label{tbl:DBPEDIA}
}
\end{figure*}

\godown

Below we report the number  returned resources,
for various entities
and for various queries,
including the query that returns the union
of
$X_{rf}(A)$ and $X_{cl}(A)$,
denoted by $X_{rf \cup cl}(A)$, defined as:
{\small
\begin{verbatim}
SELECT  ?y
WHERE{   A ?p ?x.
        ?y ?p ?x.
        FILTER (?p1 != rdf:type &&
                ?p2 != rdf:type)
     } UNION {
         A rdf:type ?z.
        ?y rdf:type ?z.
     }
\end{verbatim}
}
We did not manage to obtain reliable results
for the above queries over the LinkedMDB SPARQL endpoint,
since
for some reason it does not return very big
answers.
Therefore
at Table \ref{tbl:Reversals}
we report  some
indicative (and quite predictable) results
over the local KB.
Even in this toy KB
we can see
how the resources are reduced while the required time does not increase a lot.

To get more realistic results,
we  tried the SPARQL endpoint of DBPEDIA\footnote{http://dbpedia.org/sparql}.
If  $A$ is the movie {\tt Americano}\footnote{
        http://dbpedia.org/resource/The\_Americano
    }
then
\bequa
|X_{rf \cup cl}(A)| & = & 1,679,605 \\
|X_{rf}(A)|  & = &  32,318 \\
|X_{rf'}(A)|  & = &  32,094 \\
|X_{cl}(A)| & = & 1,679,318  \mybox{(i.e. all films)} \\
|X_{rf \cap cl}(A) & = & 32,031
\eequa
Measurements for other entities
are shown at
Table \ref{tbl:DBPEDIA}.
We observe
some
 big reductions in the answer set
(from millions to tens of thousands).
However,  even
for the  intersection query
the returned answer is quite big;
32 thousands hits
although much less than millions,
are probably
many
for fast real-time interaction.
One approach to tackle this problem
is to try formulating even more restrictive queries
which capture the desired characteristic of similarity function
in a more accurate way.
The extra condition(s) can be
added to the query as extra graph pattern,
or
the query can be  enriched
with an appropriate {\tt order by} clause.
\outcomment{
    {\bf
        Auto einai ena zhthma pou thelei skepsh.
        Any idea?
    }
}
In the latter case
the application can consume only the top hits of the
ranked hits of the computed answer.

The general approach would be to enrich the query with
 aggregated counts
 or similarity functions
aiming at reaching a query that directly returns ranked the top-$L$ similar entities.
However this is not always possible
(depends on how the similarity metric is defined),
and in some times this approach is expected to be less efficient
than getting through queries  the information that is
needed and then rank the entities using programming language code.
Of course the availability of LOD SPARQL endpoints
which support
extended versions of SPARQL
would be useful.
For instance,
\cite{kiefer2007fundamentals} investigates methods
to integrate customized similarity functions
into SPARQL.
Among the proposed techniques,
it seems that the,
so called {\em virtual triple} approach,
would be beneficial
(shorter queries which are easier to write,
optimization potential).
However, the scenarios
described
are more simple
in the sense that only
on the direct neighborhood of the compared entities
is taken into account,
and similarity thresholds should be adopted
(instead of a parameter $L$).
This direction should be further researched.
In general,
there is a need for
semantic query optimization techniques
for similarity queries.

Another important point,
which is independent of the query language,
is that the refinement of the information that is available in the LOD cloud,
i.e. the classification of the available resources
to more {\em refined} classes,
is expected to improve not only the quality of the computed similarities,
but will make the computation of the similar entities more efficient.
Specifically, if entity $A$ were not classified only as {\tt film},
but to more refined classes (e.g. {\tt Thriller}, {\tt Anti-war Film} etc),
then $|X_{cl}(A)|$ would be smaller.

\godown

Above we have sketched  a top-$L$ version of the algorithm
and identified evaluation approaches and difficulties,
for the case $k=1$.
If $k$ is greater than one,
then one approach is to start from a $k'=1$ and apply the above algorithm.
If the fetched elements are less than $L$ then move to $k'=2$,
and so on,
until
having fetched $L$ elements or reached the original value of $k$
(i.e. until $k'=k$).
However, as we saw
in the example of Figure \ref{fig:NewEx}(III),
such an approach
 does not guarantee
that the top-$L$ similar
entities
with respect to  $sim_1$
are the same with
the top-$L$ similar with respect to $sim_k$
(nevertheless  this approach could be used as an approximation).

\comment{ ===
    \godown
    {\bf
    The derivation of queries
    (combining several of the aforementioned ideas)
    which can be evaluated efficiently,
    is subject of further research.
    }
==}

\godown

Probably,
the best feasible solution, for the time being,
is to define, store and periodically update,
{\em materialized views}
accessible through LOD endpoints,
which for each entity
contain the set of
most similar entities.


\section{Conclusion}
\label{sec:Concl}

In this paper,
we motivated the need for
similarity-based browsing
over entities which are semantically defined.
This kind of browsing
can be applied
for
various kinds of entities
e.g. for
movies, paintings, photographs, videos,
restaurants,   or even social entities (groups or individual persons).
We  introduced
a similarity metric
which is  type-independent,
meaning that it can find similarities between entities of different
type (for example similarities between an actor and a movie),
which is very convenient for similarity-based browsing.
The way the similarity metric functions
is somehow similar
with the
{\em spreading activation} retrieval method
proposed for semantic networks \cite{Cohen87}.
The metric
can also be configured
(the radius $k$ as well as the graph expansion policy)
according to the characteristics
of the corpus at hand
(and the "affordable" computational complexity).
We demonstrated the behavior and the benefits
of this metric over a LOD-based  application offering
similarity-based browsing for movie information.
We  believe that this metric can also
be useful in semantic search
\cite{fazzinga2010semantic}.
We do not argue that the graph expansion method
adopted by the similarity function
is the best for all occasions.
Instead we have the impression
that in many cases
the selection of the graph expansion
method should be  application specific.

Finally, we
discussed implementation approaches
and we elaborated on
a method which is
"harmonized" with the
distributed and open nature of LOD.
The described method
can be used for computing the $L$ most similar entities
according to similarity metrics which are neighborhood-based.
Specifically we showed how a neighborhood-based similarity metric
can be reversed
to
get a query which can collect
only those entities
whose similarity is certainly greater than 0.
Furthermore we sketched
possible top-$L$
extensions of the algorithm.

Below we discuss some directions which according to our opinion are worth further research.
Regarding similarity functions
there is a need for test collections
appropriate for comparative evaluation.
Regarding algorithms,
it is worth
investigating
top-$K$ (or nearest $K$) algorithms appropriate for the LOD domain.
Regarding
services for end users,
a next step
is to device methods
for clustering the set of similar entities.
Finally, as in web searching,
log analysis can be exploited
for improving the computation of similarities
at application layer.

Moreover, we would like to note
that as the number of sources increases,
the need for ontology matching techniques
(and lexical similarity functions) increases as well.
In our application,
and since we used two sources of information,
we did not face this problem.
In any case, the approach
presented in this paper
can be applied after applying entity matching approaches.
A related issue is the management of the {\tt sameAs} predicate.
In brief,
if two entities are related with such relationships,
then they should be treated as equal by the similarity function.
Another direction is to consider weighted triples,
e.g. investigate a representation framework like Fuzzy RDF \cite{Straccia09},
and investigate similarity functions for such KBs
(an extension of the faceted browsing for such sources
is described at \cite{TzitzikasESWC11}.

\bibliographystyle{plain}
\bibliography{bib/autocomp,bib/survey,bib/diffBibBnodes}

\end{document}